\documentclass[12pt,a4paper]{article}
\usepackage{amssymb}
\usepackage{amsmath, epsfig}
\usepackage{units}

\newcommand{\be}{\begin{equation}}
\newcommand{\en}{\end{equation}}

\renewcommand{\vec}[1]{\boldsymbol{#1}}

\def\halft{{\textstyle\frac{1}{2}}}

\def\eightt{{\textstyle\frac{1}{8}}}

\title{Onset of Non-Linearity \\in the Elastic Bending of Blocks}

\author{
 M. Destrade$^{a \star }$, M.D. Gilchrist$^a$ and J.G. Murphy$^b$\\[0.8cm]
$^a$School of Electrical, Electronic, and Mechanical Engineering, \\
University College Dublin, Belfield, Dublin 4, Ireland;\\[0.5cm]
$^b$Department of Mechanical Engineering, \\
Dublin City University, Glasnevin, Dublin 9, Ireland.}

\date{$^\star$ Corresponding author. \\ email: michel.destrade@ucd.ie, phone: +353-1-716-1743}
\begin{document}

\numberwithin{equation}{section}

\maketitle

%%%%%%%%%%%%%%%%%%

\begin{abstract}

The classical flexure problem of non-linear incompressible elasticity is revisited assuming that the bending angle suffered by the block is specified instead of the usual applied moment. 
The general moment-bending angle relationship is then obtained and is shown to be dependent on only one non-dimensional parameter: the product of the aspect ratio of the block and the bending angle. 
A Maclaurin series expansion in this parameter is then found. The first-order term is proportional to $\mu$, the shear modulus of linear elasticity; the second-order term is identically zero, because the moment is an odd function of the angle; and the third-order term is proportional to $\mu(4\beta -1)$, where $\beta$ is the non-linear shear coefficient, involving third-order and fourth-order elasticity constants. 
It follows that bending experiments provide an alternative way of estimating this coefficient, and the results of one such experiment are presented.
In passing, the coefficients of Rivlin's expansion in exact non-linear elasticity are connected to those of Landau in weakly (fourth-order) non-linear elasticity.

\end{abstract}

%%%%%%%%%%%%%%%%%%%

\noindent{\bf Key words}: Incompressible elasticity; flexure; non-linear shear coefficient; experimental data.

\newpage

%%%%%%%%%%%%%%

\section{Introduction}

%%%%%%%%%%%%%%

The problem of flexure is one of the now classical problems of the theory of non-linear, incompressible elasticity. First formulated and solved by Rivlin \cite{Rivl49}, it has since been studied extensively in the literature, see for example Green and Zerna \cite{GrZe54} and Ogden \cite{Ogde84}. 
There is continuing theoretical interest in this problem, as can be seen, for example, in the recent study of Kanner and Horgan \cite{KaHo08}. 
The physical problem considered is easily visualized: a rectangular specimen is bent by equal and opposite terminal couples applied on the end faces of a rectangular block, while its other faces remain free of traction. 
Rivlin  \cite{Rivl49} shows that if a circular, annular sector is assumed for the deformed configuration, then an elegant solution to the corresponding boundary value problem can be found. 
We recall this derivation in Sections \ref{Large_plane_strain_bending} and \ref{The_stress-free_boundary_conditions}.

Typically, the usual formulation of this boundary value problem implicitly assumes that the terminal moments are specified. 
It is shown here that specifying instead the \emph{bending angle} through which the block is bent results in a simpler mathematical formulation and solution of the problem. 
We show that this solution depends on only one non-dimensional parameter: the product of the aspect ratio of the block and the bending angle, which we denote by $\epsilon$. 
If $\epsilon$  is assumed small, then the solution at low orders has a particularly simple form. 
Because $\epsilon$ is the product of the aspect ratio of the block and the bending angle, the lower-order solutions are applicable in at least two distinct physical regimes: the first corresponds to the bending of bars through an infinitesimal bending angle and the second corresponds to the non-linear bending of thin sheets. 
Although these two sub-cases are the most important, there are other possibilities: moderately thick blocks could also be considered turned through moderate angles, provided the product $\epsilon$  is small. 

\begin{figure}
\center
\epsfig{figure=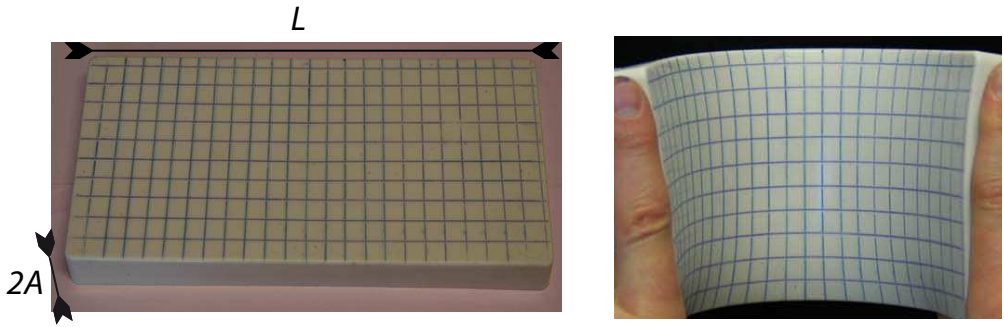, width=.99\textwidth}
 \caption{Bending of a block of silicone rubber, with length $L = 24$ cm and thickness $2A = 2$ cm.
 The dimension of each square drawn on the top surface is 1 cm $\times$ 1 cm.
 The picture on the right shows that even for such a ``home-made'' bending experiment, there exists a region about 3 cm wide around the median where plane strain is respected.}
 \label{figure1}
\end{figure}

To analyze the problem, \emph{plane strain} conditions can be assumed. 
Hence the technical interpretation of our results is that they describe the bending of planar sections of infinitely wide blocks. 
We assume that the plane strain results obtained are also applicable to strips with a finite, out of the plane dimension larger than the block thickness, i.e. that the edge/anticlastic effects are negligible. 
This assumption is motivated by observations and measurements for the bending of rubber blocks, see Gent and Cho \cite{GeCh99}.
Of course, this assumption is a limitation of Rivlin's solution, because secondary fields are bound to be observed outside of a central area in the bent block, see Figure \ref{figure1}. Nonetheless, it must be kept in mind that Rivlin's solution is one of the very few \emph{universal} solutions, valid in principle for every incompressible isotropic material, whatever the actual dimensions of the block and the amount of bending. 
For more refined deformations, albeit limited to certain types of geometries, the reader is referred to the advances obtained by Shield \cite{Shie92}. 
The other limitation of this solution is that it might bifurcate, see \cite{Haug99, CoDe08, DeNC09, DGMM10} for an in-depth treatment of this possibility. 

Given that here, the deformation depends only on the bending angle, the corresponding stress distribution can be easily determined. The most important functional of this stress distribution is the moment that needs to be applied at the ends of the block to effect the deformation. 
In Sections \ref{Some_approximations_of_the_deformed_configuration} and \ref{Exact_results_for_the_moment}, we show that the general relation between applied moment and the parameter $\epsilon$, the main relation of interest in this problem, can be expressed in a succinct and elegant form. 
It is shown that this relation is an odd function in $\epsilon$, which agrees with an intuitive expectation that the moment should be an odd function of the angle since the moment required to bend a block by an angle $\alpha$ say, is the opposite of the moment required to bend it by an angle $-\alpha$.  

On expanding the moment in a Maclaurin series in $\epsilon$ (Section \ref{Approximate_results_for_the_moment}), we find that the first-order coefficient is proportional to $\mu$, the shear modulus of linear elasticity. The second-order coefficient is identically zero for all elastic materials, because the moment is an odd function in $\epsilon$.  This suggests that the linearized moment-angle relation is likely to be valid for $\epsilon$ values beyond the infinitesimal range, and this is verified numerically and experimentally in Section \ref{Experimental_results}.
Prior to this, we show in Section \ref{Weak_non-linear_elasticity} that the third-order term in the expansion is proportional to $\mu(4\beta -1)$, where $\beta$ is the non-linear coefficient of plane shear waves \cite{ZIHM04, DeSa06, DeSa08, JCGB07}. 
Explicitly, $2\beta = (\mu + \mathcal{A}/2 + \mathcal{D})/\mu$, where $\mathcal{A}$ and $\mathcal{D}$ are Landau third- and fourth-order  elastic constants \cite{HaIZ04b}.
This coefficient has been measured for agar-based gels, based on the measurement of shear wave speeds in   transient elastography \cite{JCGB07}, or on the measurement of homogeneous plane strain deformations  \cite{ESEO04}.
Clearly, the information collected from the bending of a block, such as that provided by a bending stiffness tester \cite{ASTM}, yields a simple and useful alternative to these protocols.

In Section \ref{Experimental_results}, our main results are then compared to experimental data obtained by performing a bending experiment on a polyurethane elastomer. 
We find that $\beta \simeq 1.0$.

%%%%%%%%%%%%%%

\section{Large plane strain bending}
\label{Large_plane_strain_bending}

%%%%%%%%%%%%%%

\begin{figure}
\center
\epsfig{figure=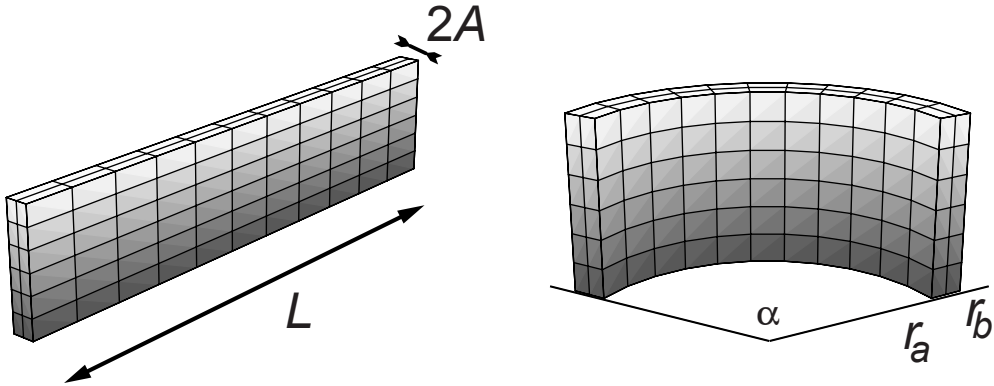, width=.99\textwidth}
 \caption{Sketch of Rivlin's deformation for the bending of a block made of an incompressible isotropic solid, with length $L$ and thickness $2A$, into a circular annular sector with inner and outer radii $r_a$ and $r_b$, respectively. The angle $\alpha$ is the \emph{bending angle}. The deformation is plane strain, but makes no assumption about the dimensions of the block or the amount of bending.}
 \label{figure2}
\end{figure}

The fundamental assumption introduced by Rivlin  \cite{Rivl49} to model the non-linear flexure of an incompressible block is that a block of length $L$ and thickness $2A$ is deformed under applied terminal moments into a circular, annular sector. For definiteness, assume that that the faces $X=\pm A$ are deformed into the inner and outer radii, denoted by $r_a$, $r_b$ respectively, of the annular sector and the faces $Y=\pm L/2$ are deformed into the faces $\theta=\pm \alpha$, where $\alpha$, the bending angle, is a $\emph{specified}$ constant. Plane strain conditions  are assumed throughout.
The \emph{bending angle}, $\alpha$, is restricted to lie in the range 
\begin{equation}
0 \leq \alpha \le \pi,
\end{equation}
which only allows a block to be bent into at most a circular annulus, see Figure \ref{figure2}.

Adopting the semi-inverse approach of Rivlin  \cite{Rivl49}, assume that 
\begin{equation}
r = r(X), \qquad \theta = \theta(Y), \qquad z = Z,
\end{equation} 
where ($X, Y, Z$) and ($r, \theta, z$) denote the Cartesian and cylindrical polar coordinates of a typical particle before and after deformation, respectively. 
Incompressibility then yields
\begin{equation}
r^2 = 2 B X + D, \qquad \theta = Y/B + C, \qquad z = Z,
\end{equation}
where $B$, $C$, $D$, are constants.

As noted by Rivlin  \cite{Rivl49}, symmetric boundary value problems can be considered without loss of generality and therefore $C \equiv 0$.
The key element in our solution of the bending problem is that the bending angle, $\alpha$, is specified and not the applied moment, as is usually assumed in most treatments, even if this assumption is implicit. 
It therefore follows easily that $B$ can be \textit{determined} as
\begin{equation}
B = L/\alpha.
\end{equation}
This yields the non-homogeneous deformation field
\begin{equation} \label{rivlin_soln}
r = \sqrt{2 (L/\alpha)X + D}, \qquad \theta = (\alpha/L)Y, \qquad z = Z,
\end{equation}
where $D$ remains to be determined.
Therefore, the inner and outer radii of the deformed curved surfaces are determined by
\begin{equation}
r _{a,b} = \sqrt{D \mp 2 (L/\alpha)A }.
\end{equation}
Adding and subtracting these equations then yields
\begin{equation}\label{r_ab}
D = (r_a^2 + r_b^2)/2, \qquad
r_b^2 - r_a^2 = 4 A L / \alpha.
\end{equation}

Hereafter we consider the boundary value problem where equal and opposite moments are applied to the ends of the block at $Y = \pm L/2$. 
The other classical boundary value problem of flexure, where one end is held fixed and a moment applied to the free end, is a subregion of the problem considered here with the fixed end described by  $\theta = 0$.

%%%%%%%%%%%%%%

\section{The stress-free boundary conditions}
\label{The_stress-free_boundary_conditions}

%%%%%%%%%%%%%%

The corresponding deformation gradient tensor, $\vec{F}$, is given by
\begin{equation} \label{F}
\vec{F} = \text{diag}\left(\lambda_1, \lambda_2, \lambda_3\right)
= \text{diag}\left(\lambda, \lambda^{-1}, 1\right), \quad \text{where}
\quad \lambda = L/(\alpha r),
\end{equation}
denoting the principal stretches by $\lambda_1$, $\lambda_2$, $\lambda_3$.
It is now clear that Rivlin's solution \eqref{rivlin_soln} is a \emph{plane strain} deformation, because $\lambda_3=1$ at all times.

For homogeneous, incompressible, elastic materials, the corresponding principal Cauchy stresses are given by
\begin{equation}
T_{rr} = -p + \lambda_1 W_{,1} , \qquad T_{\theta \theta} = -p + \lambda_2 W_{,2},
\end{equation}
where $p$ is an arbitrary scalar field,  $W =   W(\lambda_1, \lambda_2, \lambda_3)$ is the strain-energy function and the comma subscript denotes partial differentiation with respect to the appropriate principal stretch. 
The equations of equilibrium determine $p$ as
\begin{equation}
p = \int \left(\lambda_1 W_{,1} - \lambda_2 W_{,2}\right) r^{-1} \text{d}r
 + \lambda_1 W_{,1} + K,
\end{equation}
where $K$ is an arbitrary constant. 
It therefore follows immediately that
\begin{equation}
T_{rr} =  \int \left(\lambda_1 W_{,1} - \lambda_2 W_{,2}\right) r^{-1} \text{d}r
 + K,
 \qquad
T_{\theta \theta} =  T_{rr} +  \lambda_2 W_{,2} - \lambda_1 W_{,1}.
\end{equation}

Now define the function $\widetilde W(\lambda)$ as
\begin{equation}
\widetilde W (\lambda) = W(\lambda, \lambda^{-1}, 1),
\end{equation}
which is assumed to be a convex function. Then 
\begin{equation} \label{zerostress}
\lambda \widetilde{W}' = \lambda_1 W_{,1} - \lambda_2 W_{,2},
\end{equation} 
where the prime denotes differentiation.
The stress distribution can then be written simply as functions of $\lambda$  as
\begin{equation} \label{T}
T_{rr} = \widetilde W + K, \qquad T_{\theta \theta} = \widetilde W - \lambda \widetilde{W}' + K.
\end{equation}
The curved surfaces of the bent block are assumed to be free of traction.
This assumption then yields
\begin{equation} \label{K}
K = -\widetilde W (\lambda_a), \qquad \widetilde W (\lambda_b) = \widetilde W(\lambda_a),
\end{equation}
where 
\begin{equation} \label{l_ab}
\lambda_a = L/(\alpha r_a), \qquad \lambda_b = L/(\alpha r_b).
\end{equation}

No assumptions have been made thus far about material symmetry. 
Only isotropic materials will be considered here. 
For these materials, $W(\lambda_1, \lambda_2, 1) = W(\lambda_2, \lambda_1, 1)$ ,  and so \eqref{K} yields
\begin{multline} \label{sym}
W\left(\dfrac{L}{\alpha r_a}, \dfrac{\alpha r_a}{L}, 1\right) = 
W\left(\dfrac{\alpha r_a}{L}, \dfrac{L}{\alpha r_a},1\right) \\
= 
W\left(\dfrac{L}{\alpha r_b}, \dfrac{\alpha r_b}{L}, 1\right) = 
W\left(\dfrac{\alpha r_b}{L}, \dfrac{L}{\alpha r_b}, 1\right).
\end{multline} 
There are two obvious solutions to these equations:  $r_a = r_b$, which is physically unacceptable, and \cite{Rivl49}
\begin{equation} \label{sufficient}
\alpha^2 r_a r_b = L^2,
\end{equation}
which is assumed henceforth. 
Solving for $r_b$ and substitution into \eqref{r_ab}$_2$ yields a quadratic equation for  $r_a^2$, with the following unique physically acceptable solution:
\begin{equation} \label{r_a^2}
r_a^2 = \dfrac{L}{\alpha} \left( \sqrt{4 A^2 + \dfrac{L^2}{\alpha^2}} - 2A \right),
\end{equation}
which completely determines the deformed configuration. 
Unusually, it is \emph{independent} of the form of the strain-energy function (provided that \eqref{sufficient}, which is sufficient for \eqref{sym} to be satisfied, is also necessary). Also note that for isotropic materials, it follows immediately from \eqref{zerostress} that
\begin{equation} \label{zeroderiv}
\widetilde{W}'(1)=0.
\end{equation}

Substitution of \eqref{sufficient} and \eqref{r_a^2} into \eqref{l_ab} yields the following form for the stretches in terms of the non-dimensional parameter $\epsilon$:
\begin{equation} \label{lambda_ab}
\lambda_a = \sqrt{\sqrt{1+\epsilon^2} +  \epsilon}, \qquad
\lambda_b = \sqrt{\sqrt{1+\epsilon^2} -  \epsilon} = \lambda_a^{-1}, 
\end{equation}
where $\epsilon$ is the product of the block aspect ratio by the bending angle,
\begin{equation} \label{epsilon}
\epsilon = \dfrac{2 A}{L} \alpha.
\end{equation}

Finally, the qualitative features of the stress distribution will be determined. Equation \eqref{T}$_1$ yields
\begin{equation} \label{Tdiff}
\frac{\text{d} T_{r r}}{\text{d} r}=-\frac{1}{r}\lambda \widetilde{W}'(\lambda),
\qquad
\frac{\text{d}^2T_{r r}}{\text{d}r^2}=\frac{1}{r^2}\left(2\lambda\widetilde{W}'(\lambda)+\lambda^2{W}''(\lambda)\right).
\end{equation}
It follows immediately from these, the assumed convexity of $\widetilde W$, and \eqref{zeroderiv} that the radial stress has a unique minimum value of $-\widetilde W(\lambda_{a})$ at $r=L/\alpha$ and since the curved surfaces are assumed stress-free, the radial stress is therefore compressive in the interior of the bent block.

Convexity also yields that the hoop stress is a monotonically increasing function of $r$, compressive on the inner curved surface and tensile on the outer.

%%%%%%%%%%%%%%

\section{Some approximations of the deformed configuration}
\label{Some_approximations_of_the_deformed_configuration}

%%%%%%%%%%%%%%

Since the deformed configuration is independent of the form of the strain-energy function, asymptotic expansions in $\epsilon$ are valid for all incompressible elastic materials. 
Expanding \eqref{r_a^2} and its counterpart for $r_b^2$ in a Maclaurin series in $\epsilon$  yields
\begin{equation} \label{series1}
r_a = \dfrac{L}{\alpha}\left[1 - \halft \epsilon +  \eightt \epsilon^2 + \mathcal{O}(\epsilon^3)\right], \qquad
r_b = \dfrac{L}{\alpha}\left[1 + \halft \epsilon + \eightt \epsilon^2 + \mathcal{O}(\epsilon^3)\right].
\end{equation}
To the leading order, we have
\begin{equation}
r_a = r_b = L/\alpha,
\end{equation}
giving a curvature 
\begin{equation}
\kappa = \alpha/L.
\end{equation}
Thus infinitesimally thin sheets, strips, and wires are bent into a circular arc whose radius is inversely proportional to the bending angle. 

Truncating the series after the linear terms in \eqref{series1}, and thus we are now considering thin, but not infinitesimally thin, sheets, strips, and wires, yields
\begin{equation}
r_a = L/\alpha - A, \qquad r_b = L/\alpha + A,
\end{equation}
which are again remarkably simple and which tell us that the thickness of the deformed block is still $2A$. 

Retaining the \emph{quadratic} terms in the expansions  \eqref{series1} yields that, again, $r_b - r_a = 2A$.

%%%%%%%%%%%%%%

\section{Exact results for the moment}
\label{Exact_results_for_the_moment}

%%%%%%%%%%%%%%

The associated stress distribution naturally requires specification of the form of the strain-energy function in order to be determined. The most important functional of this stress distribution is the moment, $M$, required to bend the block by an angle $\alpha$. 
This moment is given by 
\begin{equation}
M = \int_{r_a}^{r_b}T_{\theta \theta} r \text{d}r,
\end{equation}
which can be re-written in terms of the parameter $\epsilon$  defined in \eqref{epsilon} as follows:
\begin{equation} \label{M}
\dfrac{M}{4A^2}=  \epsilon^{-2}\int_{\lambda_a}^{\lambda_b}\widetilde{W}(\lambda)\lambda^{-3}\text{d}\lambda + \epsilon^{-1}\widetilde{W}(\lambda_a),
\end{equation}
where $\lambda_a$ and $\lambda_b$ have been defined in terms of  $\epsilon$ through \eqref{lambda_ab}.
To derive this expression, we made use of \eqref{r_ab}$_2$, \eqref{F}$_2$, \eqref{T}, \eqref{K}$_1$, and \eqref{epsilon}.
Note that in  \cite[p.293]{Ogde84}, the last term of this expression is missing; note also that the expression of $M$ in terms of an integral in $r$ can be found in Rivlin's original paper \cite{Rivl49}, see also Kanner and Horgan  \cite{KaHo08}. 
It follows trivially from \eqref{K}$_2$ and \eqref{lambda_ab} that $M$ is an \emph{odd} function of $\epsilon$.

We remark that in this paper, $M$ is the applied moment per unit width of the block: if the block's width is $H$, then the total applied moment is $H M$.

The integral term in \eqref{M} surprisingly means that explicit relations between the moment and bending angle in terms of elementary functions are difficult to obtain in general. 
Progress can be made however for some forms of $W$. 
For example, Kanner and Horgan  \cite{KaHo08} compute $M$ for the Mooney-Rivlin material and for the Gent \cite{Gent96} material.

We focus on the following Rivlin expansion of the strain-energy density in the principal invariants of the Cauchy-Green strain tensors \cite{Rivl48},
\begin{multline} \label{Riv}
W = C_{01}(I - 3) + C_{10}(II - 3) \\+ C_{02}(I - 3)^2 + C_{11}(I - 3)(II - 3) + C_{20}(II - 3)^2,
\end{multline} 
where the $C_{ij}$ ($i+j = 1,2$) are constants to be determined from experiments (see Erkamp et al. \cite{ESEO04} for example), and 
\begin{equation}
I = \lambda_1^2 + \lambda_2^2 + \lambda_3^2, \qquad
II = \lambda_1^2 \lambda_2^2 + \lambda_2^2 \lambda_3^2 + \lambda_1^2 \lambda_3^2.
\end{equation}
This strain energy density is readily implemented in most finite element analysis packages.
It includes the neo-Hookean and Mooney-Rivlin materials. 
In the case of plane strain, such as the bending problem considered here, $I=II=\lambda^2 + \lambda^{-2} + 1$, so that $W$ reduces to
\begin{equation} \label{G}
\widetilde W = \halft \mu \left[ (\lambda^2 + \lambda^{-2}-2) + \beta (\lambda^2 + \lambda^{-2}-2)^2\right],
\end{equation}
where $\mu = 2(C_{01}+C_{10}) >0$ is the infinitesimal shear modulus, and $\beta = 2(C_{02}+C_{11}+C_{20})/\mu >0$
is the non-linear shear coefficient \cite{ZIHM04, DeSa06, DeSa08}. 
Substitution into the general moment relation \eqref{M} then yields the exact, nonlinear relation
\begin{equation} \label{M_rivlin}
\dfrac{M}{2 \mu A^2} = (1-4\beta)\left[\epsilon^{-2} \ln \left(\sqrt{1+\epsilon^2}-\epsilon\right) + \epsilon^{-1}\sqrt{1+\epsilon^2}\right] + \frac{8}{3} \beta \epsilon.
\end{equation}
The last term in this expression shows that $M$ grows unbounded with $\epsilon$, except in the special Mooney-Rivlin case for which $\beta = 0$ and where the asymptotic value is $M = 2 \mu A^2$   \cite{KaHo08}.

\begin{figure}
\center
\epsfig{figure=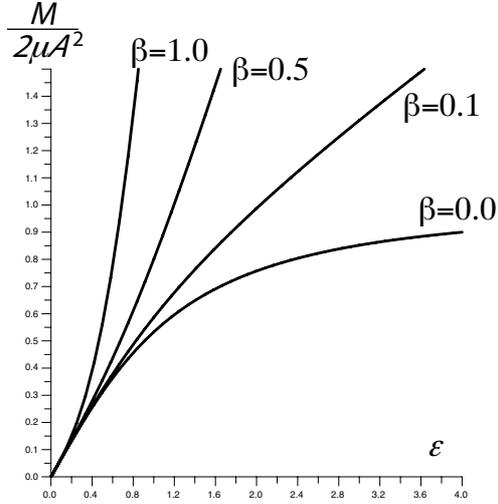, width=.47\textwidth}
 \caption{Variations of the moment $M$ with $\epsilon$, the product of the aspect ratio by the bending angle, for the large bending of an elastic block modeled by Rivlin's strain energy density \eqref{Riv}. 
Exact results, in the cases where the non-linear shear coefficient $\beta$ is equal to $0.0$ (Mooney-Rivlin material), $0.1$, $0.5$, and $1.0$.
 See Figure \ref{figure4} for a comparison with experimental results when $\beta=1.0$. }
 \label{figure3}
\end{figure}

In Figure \ref{figure3}, we plot $M/(2 \mu A^2)$ as a function of $\epsilon$ for $\beta = 0.0, 0.1, 0.5, 1.0$.
A noteworthy feature of this plot is that there is a linear response quite far beyond the origin. 
This is explained in the next section.

%%%%%%%%%%%%%%

\section{Approximate results for the moment}
\label{Approximate_results_for_the_moment}

%%%%%%%%%%%%%%

Expanding the general expression \eqref{M} for the moment $M$ in a Maclaurin series in $\epsilon$, and noting both that $M$ is odd in $\epsilon$ and \eqref{zeroderiv}, yields 

\begin{equation} \label{expan0}
\dfrac{M}{A^2} = \frac{1}{3}\left[\widetilde{W}''(1)\right]\epsilon +
 \frac{1}{120}\left[\widetilde{W}''''(1) + 8\widetilde{W}'''(1) - 3\widetilde{W}''(1)\right]\epsilon^3
+ \ldots
\end{equation}

Recall that only isotropic, incompressible materials are being considered here. 
Differentiating \eqref{zerostress} with respect to $\lambda$ and evaluation in the reference configuration yields 
\begin{equation}
\widetilde{W}''(1)=W_1+W_2+W_{11}+W_{22}-2W_{12},
\end{equation}
where the partial derivatives of $W$ are evaluated at $(\lambda_1,\lambda_2,\lambda_3)=(1,1,1)$. The last of equations (6.1.88) of Ogden \cite{Ogde84} then yields 
\begin{equation} \label{mudef}
\widetilde{W}''(1)=4\mu.
\end{equation}
The dependence of $\widetilde W$ on $\lambda$ can be expressed in the form
\begin{equation}
\widetilde W(\lambda) = f(\lambda^2+\lambda^{-2} - 2),
\end{equation}
for some function $f$, say. 
It then follows that $\widetilde{W}''(1) = 8f'(0)$ and $\widetilde{W}'''(1) = -24f'(0)  = - 3\widetilde{W}''(1)$, or 
\begin{equation}
\widetilde{W}'''(1) =-12\mu.
\end{equation}
So for isotropic materials the moment relation \eqref{expan0} can be written in the form 
\begin{equation} \label{expan}
\dfrac{M}{A^2} =   \textstyle{\frac{4}{3}}\mu\epsilon
+ \frac{1}{120}\left[\widetilde{W}''''(1) -108\mu\right]\epsilon^3
+ \mathcal{O}(\epsilon^5),
\end{equation}
which explains why the linear regime carries for moderate values of $\epsilon$ in Figure \ref{figure3}.

%%%%%%%%%%%%%%

\section{Weak non-linear elasticity}
\label{Weak_non-linear_elasticity}

%%%%%%%%%%%%%%

Because we are looking at small, but not infinitesimal, elastic effects, we place ourselves in the theory of \emph{weak non-linear elasticity}  \cite{LaLi86}.
There, the strain energy density is expanded in terms of 
\begin{equation} \label{I_i}
I_1 = \text{tr}(\vec{E}), \qquad
I_2 = \text{tr}(\vec{E}^2), \qquad
I_3 = \text{tr}(\vec{E}^3), 
\end{equation}
where $\vec{E}$ is the Green-Lagrange strain tensor (with eigenvalues $(\lambda_i^2-1)/2$).
For \emph{incompressible} solids, Ogden \cite{Ogde74} shows that the expansion of $W$ up to terms which are of order four or less in the Green-Lagrange strain involves only three material constants.
In the notation of Hamilton et al. \cite{HaIZ04b}, it is written as
\begin{equation} \label{4th}
W = \mu I_2 + \frac{\mathcal{A}}{3} I_3 + \mathcal{D} I_2^2,
\end{equation}
where $\mathcal{A}$ and $\mathcal{D}$ are nonlinear Landau elasticity constants. 

The Appendix shows that \emph{at the same order of approximation} in the strains, the Rivlin strain energy density \eqref{Riv} coincides with the fourth-order elasticity expansion \eqref{4th} when
\begin{equation} \label{connections} 
\mathcal{A} = -8(C_{01} + 2C_{10}),
\qquad
 \mathcal{D} = 2(C_{01} + 3C_{10} + 2C_{02} + 2C_{11} + 2C_{20}).
\end{equation} 
Conversely, we find that $\beta$ introduced in \eqref{G} can be written as
\begin{equation}
\beta = \frac{1}{2}\left( 1 + \frac{\mathcal{A}/2+ \mathcal{D}}{\mu}\right),
\end{equation}
in agreement with Zabolotzkya et al. \cite{ZIHM04}.

Consequently the general moment-$\epsilon$ relation \eqref{expan} for fourth-order elasticity has the form
\begin{equation} \label{expan2}
\dfrac{M}{A^2} =  \textstyle{\frac{4}{3}} \mu \epsilon
+ \textstyle{\frac{2}{5}}(\mu + \mathcal{A} + 2\mathcal{D})\epsilon^3
+ \mathcal{O}(\epsilon^5),
\end{equation}
or, in non-dimensional form,
\begin{equation} \label{expan3}
\dfrac{M}{\mu A^2} =  \textstyle{\frac{4}{3}} \epsilon
+ \textstyle{\frac{2}{5}}(4\beta -1)\epsilon^3
+ \mathcal{O}(\epsilon^5).
\end{equation}
We remark that this expansion is in agreement with the Maclaurin series derived from the exact expression \eqref{M_rivlin}, see Rivlin \cite{Rivl49}.
We also note that \emph{fourth-order elasticity is necessary and sufficient} to express the onset of non-linearity present in the coefficient of $\epsilon^3$ in \eqref{expan}: third-order elasticity cannot account for all the components of that coefficient, and we checked that fifth-order constants do not appear in it (these calculations are not reproduced here).

Finally, we check that the expansion \eqref{expan3} is consistent with classic elastic theory, which tells us that the total flexural moment required to achieve a curvature $\kappa$ of a plate with width $H$ and thickness $2A$ is
\begin{equation} \label{classic}
H M = E I \kappa,
\end{equation}
where $E$ is Young's modulus and $I$ is the moment of inertia of cross-sectional area. 
Here the block is rectangular and $I = H (2A)^3/12$.
Also, the curvature is $\kappa = \alpha/L$, see Section \ref{Some_approximations_of_the_deformed_configuration}.
Moreover, it is known that in classic \emph{plane strain} theory, $E= 2\mu (1+\nu)/(1-\nu^2)$, where $\nu$ is Poisson's ratio. 
As we are dealing only with incompressible solids, $\nu = 1/2$, and we find that the linear term in \eqref{expan3}, or in \eqref{expan} (or the term obtained from a linear expansion in $\epsilon$ of \eqref{M_rivlin}) is indeed given by \eqref{classic}.

%%%%%%%%%%%%%%

\section{Experimental results}
\label{Experimental_results}

%%%%%%%%%%%%%%

We conducted bending experiments on several strips of elastomers, using a Tinius Olsen bending stiffness tester. 
That tester meets the requirements of the ASTM standard test method E855 \cite{ASTM}.
We used strips which were about 4.5 mm thick. 
The tester imposes a moment at two points of the strip separated by half-an-inch. 
Hence the aspect ratio of the strips, with respect to the bending experiments, was $A /L = 4.5/12.7 \simeq 0.35$.
We bent the samples by small, moderate, and large bending angles, but noticed that at large angles, pinching (and perhaps also wrinkling \cite{CoDe08}) took place on the inner face of the bent strip.
Consequently we only retained the data up to an angle of 60$^\circ$, which gives the following range for the expansion parameter: $0 \leq \epsilon \leq 0.37$.

\begin{figure}
\center
\epsfig{figure=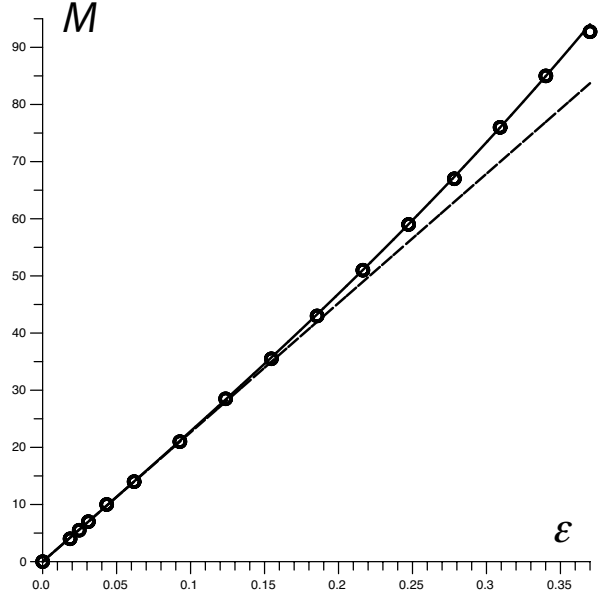, width=.75\textwidth}
 \caption{Bending of a strip of polyurethane: variation of the non-dimensional measure of the bending moment $M/M_m$ with $\epsilon$, the product of the strip's aspect ratio by the angle of bending.
 Circles: experimental data; Dashed straight line: linear elasticity, equation \eqref{classic}; Full line plot: fitting of third- and fourth-order elasticity effects with the data, by adjusting the non-linear parameter $\beta$ to the value 1.0 in \eqref{expan3}.}
 \label{figure4}
\end{figure}

In Figure \ref{figure4}, we show the results of one representative experiment, for a strip of polyurethane elastomer, shore hardness 40A. 
On the vertical axis, the variable is a non-dimensional measure of the moment, $M/M_m$, where $M_m$ is the maximum bending moment of the tester's pendulum (see ASTM standard test method E855 \cite{ASTM} for details); the actual value of $M_m$ is irrelevant, as we are only interested in measuring the non-dimensional parameter $\beta$.
On the horizontal axis, the variable is $\epsilon$.
The circles represent the recorded experimental data (16 measurements in the 0-60$^\circ$ degree range for the angle of bending).
The straight dashed line corresponds to the fitting with linear elasticity theory ($\beta=0$, equation \eqref{classic}), and the full thick plot corresponds to the fitting with fourth-order elasticity theory  \eqref{expan3}.
Only one parameter ($\beta$) is to be determined from the cubic relation \eqref{expan3}, which ensures the existence and unicity of the fitting parameter $\beta$.
We obtained a good agreement when $\beta = 1.0$.

 \newpage
 
\begin{center}
\emph{Table 1. Experimental results collected for the bending of a polyurethane strip with aspect ratio $A /L  \simeq 0.35$.
The first line gives the bending angle $\alpha$ in degrees;
the second line gives a measure of the moment $M$, up to a multiplicative factor.}
\\[8pt]
{\small
  \begin{tabular}{ l | l | l | l | l | l | l | l | l | l | l | l | l | l | l | l| }
    \hline
    $\alpha$  & 3   & 4   & 5 & 7  & 10 & 15 & 20   & 25   & 30 & 35 & 40 & 45 & 50 & 55 & 60 \\ \hline
    $\frac{M}{M_m}$       & 4 & 5${\nicefrac{1}{2}}$ & 7 & 10 & 14 & 21 & 28${\nicefrac{1}{2}}$ & 35${\nicefrac{1}{2}}$ & 43 & 51 & 59 & 67 & 76 & 85 & 93 \\
    \hline
  \end{tabular}
}
\end{center}

\bigskip

%%%%%%%%%%%%%%%%%%%%%%%%%%%%%

\section*{Acknowledgments}

%%%%%%%%%%%%%%%%%%%%%%%%%%%%%

This work is supported by a Senior Marie Curie Fellowship awarded by the Seventh Framework Programme (European Commission) to the first author.

We are most grateful to Steve Taylor (Tinius Olsen, UK) and to Gerard Keane, Alastair Goldring, and Pat McLoughlin (Abbott Vascular, Ireland) for their generous and crucial help in setting up the bending experiments.

%%%%%%%%%%%%%%

  \renewcommand{\theequation}{A-\arabic{equation}}
  % redefine the command that creates the equation no.
  \setcounter{equation}{0}  % reset counter 
  \section*{APPENDIX: Correspondence between exact non-linear elasticity and weakly non-linear elasticity}

%%%%%%%%%%%%%%%%%%%%%%%%%%%%

In the exact (finite) theory of non-linear elasticity, there are no restrictions to impose on the magnitude of the strain. 
Often the strain-energy density $W$ is written in terms of the first three principal invariants of the Cauchy-Green right strain tensor $\vec{C}$,
\begin{equation}
I = \text{tr}(\vec{C}), \qquad
II= \halft [(\text{tr} \vec{C})^2 - \text{tr}(\vec{C}^2)], \qquad
III = \det \vec{C}.
\end{equation}
For incompressible solids, $III=1$ at all times, and $W=W(I,II)$ only.

In the weakly non-linear theory of elasticity, $W$ is expanded up to a certain order in a certain measure of strain, and all higher order terms are neglected. 
Often the Green-Lagrange strain tensor $\vec{E} = (\vec{C}-\vec{I})/2$ is favoured, and the expansion is made in terms of the quantities $I_1$, $I_2$, $I_3$ defined in \eqref{I_i}.

There exist of course connections between the two theories. 
For instance, Rivlin and Saunders \cite{RiSa51} show that the Mooney strain-energy density of exact non-linear incompressible elasticity,
\begin{equation}
W = C_{01}(I-3) + C_{10} (II-3),
\end{equation}
coincides, at the same order of approximation, with the general weakly non-linear third-order elasticity expansion,
\begin{equation}
W = \mu I_2  + \frac{\mathcal{A}}{3} I_3.
\end{equation}
The connections between the material constants $C_{01}$, $C_{10}$, $\mu$, and $\mathcal{A}$ are
\begin{equation}
\mu = 2(C_{01} + C_{10}), \qquad 
\mathcal{A} = -8(C_{01} + 2C_{10}), 
\end{equation} 
or conversely, 
\begin{equation}
C_{01} = \frac{1}{2}\left(2\mu+ \frac{\mathcal{A}}{4}\right), \qquad 
C_{10} = -\frac{1}{2}\left(\mu+ \frac{\mathcal{A}}{4}\right),
\end{equation}
see Goriely et al. \cite{GoVD08}, where there is a misprint.
Now we show how Rivlin's strain-energy density \eqref{Riv} is connected to the fourth-order elasticity expansion \eqref{4th} for incompressible solids.

The general relations between $I$, $II$, $III$, and $I_1$, $I_2$, $I_3$ are well-known and straight-forward to derive:
\begin{align} \label{A6}
 I & = 3 + 2 I_1, \notag \\
 II & = 3 + 4I_1 - 2I_2 + 2I_1^2, \notag \\
 III & = 1 + 2I_1 + 2I_1^2 - 2I_2 + \textstyle{\frac{4}{3}} I_1^3 - 4I_1I_2 + \textstyle{\frac{8}{3}} I_3.
\end{align}
For incompressible solids, $III=1$ is enforced at all times, and so
\begin{equation} \label{A7}
I_1 = -I_1^2 +I_2 - \textstyle{\frac{2}{3}} I_1^3 + 2I_1I_2 - \textstyle{\frac{4}{3}} I_3,
\end{equation}
showing that $I_1$ is at least of order 2.
Squaring gives
\begin{equation} \label{A8}
I_1^2 = I_2^2 + \text{H.O.T.},
\end{equation}
where ``H.O.T.'' is the acronym for ``Higher Order Terms'' (here, higher than fourth-order terms).
Multiplying \eqref{A7} by $I_2$ yields $I_1I_2 = I_2^2 - I_1^2I_2 + \text{H.O.T.}$ or, using \eqref{A8},
\begin{equation} \label{A9}
I_1I_2 = I_2^2 + \text{H.O.T.}
\end{equation}
Substituting \eqref{A8} and \eqref{A9} into \eqref{A7} gives \cite{HaIZ04b, JCGB07}
\begin{equation}
I_1 = I_2 - \textstyle{\frac{4}{3}} I_3  + I_2^2  + \text{H.O.T.}.
\end{equation}
Hence, the relations \eqref{A6} reduce, for incompressible solids, to
\begin{align} \label{A10}
 & I -3  =   2 I_2 - \textstyle{\frac{8}{3}} I_3  + 2I_2^2  + \text{H.O.T.}, \notag \\
 & II -3 = 2I_2 - \textstyle{\frac{16}{3}} I_3  + 6 I_2^2  + \text{H.O.T.}, \notag \\
 & (I -3)^2 = (I-3)(II-3) = (II-3)^2 = 4 I_2^2  + \text{H.O.T.},
\end{align}
and Rivlin's strain-energy density \eqref{Riv} reduces to 
\begin{equation}
W = 2(C_{01} + C_{10})I_2 - \frac{8}{3} (C_{01} + C_{10})I_3 
 + 2(C_{01} + 3C_{10} + 2C_{20}+ 2 C_{11}+ 2 C_{02})I_2^2.
\end{equation}
Clearly, it coincides with the fourth-order elasticity expansion \eqref{4th}, with the connections \eqref{connections}.

%%%%%%%%%%%%%%%%%%%%%%%%%%%%

%%%%%%%%%%%%%%%%%%%%%

\end{document}